\documentclass[aps]{revtex4}
\usepackage{epsfig,amsmath,graphicx,amssymb}
\usepackage[colorlinks=fals,linkcolor=blue]{hyperref}

\def\be{\begin{equation}}
\def\ee{\end{equation}}
\def\bea{\begin{eqnarray}}
\def\eea{\end{eqnarray}}
\begin{document}
\title{A one-dimensional continuous model for carbon nanotubes}
\author{Xiaohua Zhou}
\email{xhzhou08@gmail.com} \affiliation{Department of Mathematics
and Physics, Fourth Military Medical University, Xi'an 710032,
China}

\date{\today}

\begin{abstract}
The two-dimensional (2D) continuous elastic energy model for
isotropic tubes is reduced to a one-dimensional (1D) curvature
elastic energy model strictly. This 1D model is in accordance with
the Kirchhoff elastic rod theory. Neglecting the in-plane strain
energy in this model, it is suitable to investigate the nature
features of carbon nanotubes (CNTs) with large deformations and can
also reduce to the string model in [Z.C. Ou-Yang \emph{et al}.,
Phys. Rev. Lett. \textbf{76} 4055 (1997)] when the deformation is
small enough. For straight chiral shapes, this general model
indicates that the difference of the chiral angle between two
equilibrium states is about $\pi/6$, which is consistent with the
lattice model. It also reveals that the helical shape has lower
energy for per atom than the straight shape has in the same
condition. By solving the corresponding equilibrium shape equations,
the helical tube solution is in good agreement with the experimental
result, and super helical shapes are obtained and we hope they can
be found in future experiments.
\end{abstract}
\pacs{61.46.Np, 62.20.D-}
 \maketitle

%%%%%%%%%%%%%%%%%%%%%%%%%%%%%%%%%%%%%%%%%%%%%%%%%%%%%
\section{Introduction}
In the past two decades, the CNTs initially synthesized by Iijima
\cite{Iijima1} attracted many researchers' attention due to the
excellent physical characteristics and potential applications in
many apparatus and nano-instruments, such as field emission sources
\cite{Heer}, probe tips \cite{Dai,Wong} and quantum wires
\cite{Tans}. Experiments indicate that the configurations of CNTs,
such as their radii, lengthes and helicity, strictly determine their
physical capability, but it is difficult to precisely control those
configurations during production processes. So, although the CNT is
a particularly important functional nanomaterial, obtaining
macroscopical bulk materials is a challenging problem
\cite{Terrones}. Recently, an important progress reported by Davis
\emph{et al.} \cite{Davis} shows a way to obtain macroscopical
fibres of the SWNTs using the self-assemble method in
chlorosulphonic acid. Moreover, the mechanical parameters of CNTs
are not unification due to the size effect, which confines their
applications as a reliable high strength material. For instance, the
Young's modulus will decline with the increase of the diameters of
CNTs \cite{Wang}. The above problems indicate that there are still
some challenges to overcome before we can widely put CNTs in
practice.

An interesting phenomenon is that CNTs often present as beelines
(including zigzag and armchair shapes \cite{Iijima2,Thess,Dekker}),
helixes \cite{ZhangXB,ZhangM1,ZhangM2} as well as rings
\cite{Martel,Cohen}. And a possible reason for those shapes is the
thermodynamic effects in different synthesizing methods.
Particularly, periodic defects (heptagon and pentagon cells) play an
important rule in the forming of helical shapes
\cite{ZhangXB,Dunlap}. SWNTs are generally taken as a frizzy
graphite layer and multi-walled carbon nanotubes (MWNTs) consist of
multiple rolled layers of graphite. Many physical properties of the
CNT are obtained by calculating the interaction between its carbon
lattices. There are also many researchers who take the CNT as 2D
continuous tubules \cite{OuYang1,Falvo,TuZC,TuZC2008}. Ou-Yang
\emph{et al.} \cite{OuYang1} continued the lattice model which
provided by Lenosky \emph{et al.} \cite{Lenosky} and pointed out
that SWNT's free energy is similar to the model for vesicles
\cite{OuYang2} when ignored the in-plane strain energy. In
Refs.~\cite{TuZC,TuZC2008} Tu and Ou-Yang provided a general 2D
model which considered the in-plane strain energy and revealed that
the effective Young's modulus of MWNTs dependents on the layer
number.

  Although the CNT is taken as the 1D elastic material, by far
there is still not a strict model to connect the lattices model and
the Kirchhoff elastic rod theory. The string model \cite{OuYang1}
provided a recommendable way to make up this missing link, but it
cannot be used to investigate the mechanical behaviors when outside
forces act on CNTs, because the in-plane strain energy can not be
ignored in this circumstance. So it needs to construct a complete 1D
model which should contain the in-plane strain energy to connect the
lattices model and the Kirchhoff theory. Besides, although the
lattice model can geometrically tell us that the defects induce the
helical shapes, it lacks a reasoned physical theory to expound the
reason of the emergence of helical CNTs. Can the continuous model
explain why there are so many helical CNTs? Further, considering
that 1D structures can form super helical shapes, such as DNA
chains, so can we obtain super helical CNTs? These questions need to
be investigated deeply. Not only will they help us to understand the
physical characters of the low dimensional systems, they will
provide us with new materials and methods to design
nano-instruments. In this paper, we will give a complete 1D
continuous CNTs model which contains the in-plane strain energy to
discuss the above problems. This paper is organized as follows: In
Sec.~\ref{perfect model}, the 2D continuous elastic shell model is
reduced to a 1D curvature elastic model strictly. This complete 1D
model is in accordance with the isotropic Kirchhoff elastic rod
model and is suitable to investigate the mechanical behaviors of
CNTs. In Sec.~\ref{defective model}, a concise model which ignores
the in-plane strain energy is used to study the nature features of
CNTs. By solving the corresponding shape equations, the helical
solution is in good agreement with the experimental result and super
helical shapes are obtained. Finally, a short discussion is
presented in Sec.~\ref{Conclusions}.
\section{A complete 1D model for isotropic elastic tubes and CNTs}
\label{perfect model}

 Let the central line of an elastica tube be
$\textbf{R}=\textbf{R}(s)$, $\vec{\alpha}=\dot{\textbf{R}}$ be the
tangent vector (an overdot denotes differential with respect to $s$
which is the arclength of the central line of the tube),
$\boldsymbol{\vec{\beta}}=\ddot{\textbf{R}}/K$ be the main normal
vector and $ \boldsymbol{\gamma}=\vec{\alpha}\times\vec{\beta}$ be
the binormal vector, between those unit vectors, there are the
Frenet formulaes: $d\vec{\alpha}/ds=K\vec{\beta}$,
$d\vec{\beta}/ds=-K\vec{\alpha}+\tau\vec{\gamma}$ and
$d\vec{\gamma}/ds=-\tau\vec{\beta}$, where $K$ and $\tau$ are the
curvature and torsion \cite{Struik} of the central line
$\textbf{R}$, respectively. The shape of the tube can be obtained by
this way: the central point of a ring with the radius $r_0$ moves
along a line $\textbf{R}$ and keeps the ring upright to the tangent
of $\textbf{R}$. It means the ring is in the normal plane of the
central line $\textbf{R}$.
 Let $\textbf{Y}$ be the shape of the tube, there is
\bea\label{Y1}
\textbf{Y}=\textbf{R}+\textbf{r}=\textbf{R}+r_0\cos\theta\vec{\beta}+r_0\sin\theta\vec{\gamma},
\eea
where the parameter $0\leq\theta\leq\pi/2$. Using the Frenet
formulaes, we obtain the mean curvature and Gaussian curvature:
$H=\frac{1-2r_0K\cos\theta}{2r_0(r_0K\cos\theta-1)}$ and
$\Lambda=\frac{K\cos\theta}{r_0(r_0K\cos\theta-1)}$, respectively.
Here we give a method in appendix \ref{appendix-A} to show how to
attain these two curvatures. The 2D curvature elastic energy of a
tube is \cite{TuZC,Landau,Yakobson1,Yakobson2,Cadelano}
\bea\label{E-S}
    F_s = \frac{Y h^3}{24(1-\nu^2)}\oint
    \left[(2H)^2-2(1-\nu)\Lambda\right]d\sigma,
\eea
where $Y$ is the Young¡¯s modulus, $\nu$ is the Poisson ratio, $h$
is the thickness of the tube and $d\sigma$ is the area element. The
above energy derives from bending the graphene to SWNTs. For any
compact, closed 2D surface, $\oint\Lambda d\sigma$ is a constant
because there is the Gauss-Bonnet theorem \cite{Carmo}
\bea\label{Er}
\oint\Lambda d\sigma =2\pi E_r,
\eea
where $E_r$ is the Euler characteristic which only depends on the
topological structure of the surface. For spherical topology
surface, there is $E_r=2$. For torus, cylinders and tubes with
infinite length, there is $E_r=0$. In the following text, we will
study the torus and tubes with infinite length, so we eliminate the
constant term associated with $\oint\Lambda d\sigma$ in (\ref{E-S}).
Considering $d\sigma=r_0(1-r_0K\cos\theta)d\theta$ and $r_0^2K^2<1$,
and using the Euler integral, Eq.~(\ref{E-S}) is reduced to
\bea\label{E-C}
    F_s=\frac{\pi Y h^3}{12r_0(1-\nu^2)}\int\frac{1}{\sqrt{1-r_0^2K^2}} ds.
\eea

\begin{figure}\centering
\includegraphics[scale =0.2]{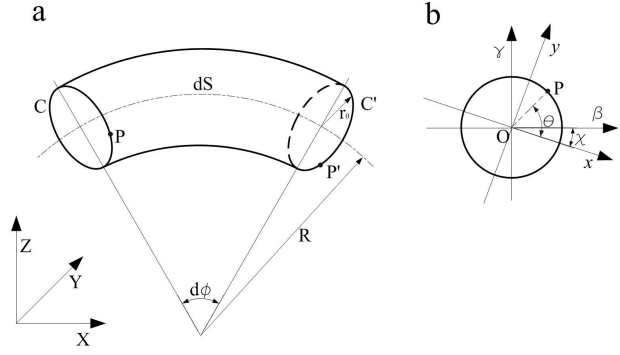}
\caption{\label{fig1} (a) A fragment of a curving SWNT with the
radius $r_0$, the length of central line $ds$ and the bending angle
$d\phi$, $C$ and $C'$ are the cross section rings on the two ends.
(b) The left cross section of the SWNT. $\vec{\beta}$ and
$\vec{\gamma}$ are the main normal vector and the binormal vector of
the central line, respectively. $(x,y,z)$ are the local coordinates
with the $x,y$ axes fixed on the cross section ring and with the $z$
axis superposed to the tangent of the central line. $\chi$ is the
angle between $\vec{\beta}$ and the $x$ axis. Point $P$ is fixed on
the cross section ring (on the left end) and the angle between
$\overline{OP}$ and $\vec{\beta}$ is $\theta$. When the cross
section ring moves along the central line from the left end to the
right end, $P$ changes to $P'$ (Note that the cross section ring can
turn around the central axis.). In this process, the angular
displacement of point $P$ in the local coordinates $(x,y,z)$ is
$d\theta$ and the change of $\chi$ is $d\chi$. Supposing that the
initial state is a straight tube, this two charts indicate the
protraction of the point $P$ along the central line is
$ds_0=-r_0\cos\theta d\phi$, the corresponding tensile strain is
$\varepsilon_x=\frac{ds_0}{ds}=-r_0K\cos\theta$ and the shear strain
is $\varepsilon_{xy}=r_0(\tau+\dot{\chi})/2$.}
\end{figure}

The in-plane strain energy can be expressed as
\cite{Landau,Yakobson1,Yakobson2,Cadelano}
\bea\label{E-I}
\nonumber   F_i =\frac{Y h}{2(1-\nu^2)}\oint
    \left[(\varepsilon_x+\varepsilon_y)^2-2(1-\nu)(\varepsilon_x\varepsilon_y-\varepsilon_{xy}^2)\right]d\sigma,\\
\eea
where $\varepsilon_x$, $\varepsilon_y$ and $\varepsilon_{xy}$ are
the axial, circumferential, and shear strains, respectively. In
Fig.~\ref{fig1}(a), we show a fragment of SWNT with the radius $r_0$
and the length of its central line is $ds$. We suppose that its
central line has a bend angle $d\phi$, so the radius of curvature is
$R=1/K=ds/d\phi$. Taking the SWNT as an isotropic elastic tube and
the central line is nonretractable, and supposing that the initial
state is a straight tube, from Fig.~\ref{fig1} we can see that the
protraction of the point $P$ along the axis direction is
$ds_0=[R-r_0\cos(\theta+d\theta)]d\phi-ds\simeq-r_0\cos\theta
d\phi$. Then the tensile strain on point $P$ is
$\varepsilon_x=\frac{ds_0}{ds}=-r_0K\cos\theta$. We simply choose
the circumferential strain $\varepsilon_y=-\nu\varepsilon_x=\nu
r_0K\cos\theta$. Correspondingly, the shear strain on point $P$ is
$\varepsilon_{xy}=\frac{1}{2}r_0\frac{d\theta}{ds}$. Note
$\frac{d\theta}{ds}=\tau+\frac{d\chi}{ds}$, there is
$\varepsilon_{xy}=r_0(\tau+\dot{\chi})/2$. Here the definitions of
$\theta$ and $\chi$ please see the caption of Fig.~\ref{fig1}. Using
the above results, Eq.~(\ref{E-I}) is reduced to
\bea
   F_i=\frac{\pi r_0^3Y h}{2(1+\nu)} \int\left[(1+\nu)K^2+(\tau+\dot{\chi})^2\right] ds.
\eea
Then, the 1D energy density of a tube can be written as
\bea\label{E-D-Shell}
 \nonumber   \mathcal{F}&=&\frac{\pi Y h^3}{12r_0(1-\nu^2)}\frac{1}{\sqrt{1-r_0^2K^2}}
                            +\frac{\pi r_0^3Y h}{2}K^2\\
                        & & +\frac{\pi r_0^3Y h}{2(1+\nu)}(\tau+\dot{\chi})^2.
\eea
For a multilayered tube, the energy density is
\bea\label{E-D-Tube-1}
    \mathcal{F}_{m}\simeq\int_{\rho_i}^{\rho_o}(\mathcal{F}/b+2\pi r_0
    g/b)dr_0,
\eea
where $b$ is the distance of two neighbor layers, $g$ is the surface
energy density between two neighbor layers, $\rho_i$ and $\rho_o$
are the inmost and outmost radii, respectively. Specially, when
$b=h$, there is
\bea\label{E-D-Tube-2}
\nonumber    \mathcal{F}_{m}&=&\frac{\pi Y
                            b^2}{12(1-\nu^2)}\ln\left[\frac{\rho_o\big(1+\sqrt{1-K^2\rho_i^2}\big)}{\rho_i(1+\sqrt{1-K^2\rho_o^2})}\right]\\
                      & &+\frac{1}{2}Y I_r K^2+\frac{1}{2}G I_t(\tau+\dot{\chi})^2+ PS_0,
\eea
where $I_r=\pi(\rho_o^4-\rho_i^4)/4$ and
$I_t=\pi(\rho_o^4-\rho_i^4)/2$ are the moments of inertia of the
cross section around its diameter and central axis, respectively.
$G=Y/(2+2\nu)$ is the shear modulus, $P=g/b$ can be taken as the
volumetric energy density and $S_0=\pi(\rho_o^2-\rho_i^2)$ is the
area of the cross section of the tube. When $\rho_i=0$, the second
and the third terms in the right hand side of the above equation
compose the typical isotropic Kirchhoff elastic rod model.

In Refs.~\cite{TuZC,TuZC2008}, Tu and Ou-Yang continued the lattice
model and showed that the energy density for SWNTs is similar to the
elastic shell model. Using their results and considering
Eq.~(\ref{E-D-Shell}), the energy density for SWNTs is
\bea\label{E-D-SWNT}
    \mathcal{F}_{sw}=\frac{A}{\sqrt{1-r_0^2K^2}}+\frac{B}{2}K^2+\frac{C}{2}(\tau+\dot{\chi})^2,
\eea
where $A=\pi k_c/r_0$, $B=(1-\nu^2)\pi r_0^3k_d$, $C=B/(1+\nu)$,
$k_c=1.62$ eV and $k_d=22.97$ eV/{\AA}$^2$ \cite{TuZC2008}.
Considering $r_0^2K^2\ll1$, $r_0=0.5\sim10$ nm and $B\gg\pi k_cr_0$,
the first term on the right hand side of the above equations can be
ignored and similar models are discussed in Refs.~\cite{Kessler,
Fonseca} to study the mechanical properties of nanosprings. The
energy density for MWNTs is
\bea\label{E-M}
\mathcal{F}_{mw}\simeq\int_{\rho_i}^{\rho_o}(\mathcal{F}_{sw}/b+2\pi
r_0 g/b)dr_0,
\eea
where $b=0.34$ nm and $g=-2.04$ eV/nm$^2$ \cite{Girifalco}.
Specially, for a straight double-walled carbon nanotube (DWNT),
there is $K=\tau=0$ in (\ref{E-D-SWNT}). Moreover, we should note
that $\chi$ is just the chiral angle in this case. Considering each
straight layer often has fixed chiral angle, there is $\dot{\chi}=0$
(actually, this result means there is no strain energy, similar
result will be shown in the later text), thus the energy density for
a DWNT can be written as
\bea\label{DWNT}
    \mathcal{F}_{dw}=\pi k_c\left(\frac{1}{r_0}+\frac{1}{r_0+b}\right)+2\pi (r_0+b/2)g,
\eea
where $r_0$ is the inner radius. Choosing $k_c=1.4$ eV
\cite{Chopra}, $\mathcal{F}_{dw}=0$ yields
\bea
    r_0=6.8~{\textrm{\AA}}.
\eea
For dense SWNT ropes, supposing that each SWNT is continuously
enveloped by its neighbors, the model in (\ref{DWNT}) can give the
optimal radius for SWNTs and the similar result obtained by Zhang
\emph{et al.} is $6.8$ {\AA} \cite{ZhangSL2}.

In the above discussions, the energy density in (\ref{E-D-SWNT})
which contains the contribution of the in-plane deformation is
suitable to study the mechanical behaviors of SWNTs. However, if we
want to study the natural features of SWNTs without any outside
forces, the in-plane deformation terms in (\ref{E-D-SWNT}) should be
neglected. The corresponding model will be studied in the following
section.

\section{1D model for CNTs without in-plane strain energy}\label{defective
model}
\subsection{ Analytical results}
\label{Analytical}

\begin{figure}\centering
\includegraphics[scale =0.2]{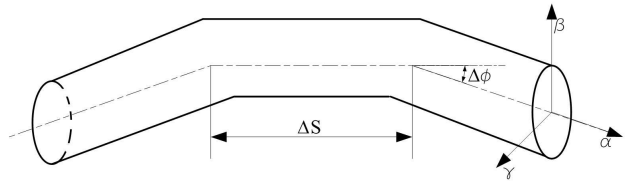}
\caption{\label{fig2} A curving SWNT is composed of several straight
segments connected by abrupt corners. The average length for each
segment is $\Delta S$ and the average corner between two neighbor
segments' central lines is $\Delta\phi$, so the curvature for the
central line is $K\simeq\Delta\phi/\Delta S$.}
\end{figure}
According to the experimental observation \cite{ZhangXB, ZhangXF},
helical SWNTs with periodic defects can be treated as the shape in
Fig.~\ref{fig2}. Seeing Fig.~\ref{fig2}, a SWNT is composed of
several straight segments connected by abrupt corners. The average
length for each segment is $\Delta S$ and the average corner between
two neighbor segments' central lines is $\Delta\phi$, so the
curvature for the central line is $K\simeq\Delta\phi/\Delta S$.
Supposing that the average torsion angle around the $\vec{\alpha}$
axis between two neighbor segments is $\Delta\theta$ (
$\Delta\theta$ here is similar to $d\theta$ in Fig.~\ref{fig1}),
there is $\tau\simeq(\Delta\theta-\Delta\chi)/\Delta S$. However,
although the central line of the shape in Fig.~\ref{fig2} has
curvature and torsion, which are due to the abrupt corners not to
the in-plane strain. Thus the second and third terms on the right
side hand of Eq.~(\ref{E-D-SWNT}) should be ignored for SWNTs with
out outside forces. Seeing an example, for a helical shape with
$K=\frac{R}{R^2+h^2}$ and $\tau=\frac{h}{R^2+h^2}$, we define
$\omega=h/R$, there is
\bea\label{H-A}
\Delta\phi=\omega(\Delta\theta-\Delta\chi),
\eea
here $\Delta\chi$ is the chiral angel difference between two
neighbor straight segments. In \cite{ZhangXB}, Zhang \emph{et al.}
found $\Delta\phi=\pi/6$, and particularly in Fig.~5a of
\cite{ZhangXB}, we can see that two neighbor segments are composed
by a zigzag shape and an armchair shape, which indicates
$\Delta\chi=-\pi/6$. Moreover, Ou-Yang \emph{et al.}\cite{OuYang1}
pointed the shape in Fig.~2a of \cite{ZhangXB} satisfies $\omega=1$,
then substituting the above results into Eq.~(\ref{H-A}) yields
\bea\label{H-B}
\Delta\theta=0.
\eea
This result indicates $\tau+\dot{\chi}\simeq\Delta\theta/\Delta S=0$
in Eq.~(\ref{E-D-SWNT}) strictly. The length for per helix turn is
$S_p=2\pi\sqrt{R^2+h^2}$, and the segments number for per helix turn
is
\bea\label{H-C}
  N=\frac{S_p}{\Delta S}=\frac{2\pi}{\Delta\phi}\frac{1}{\sqrt{1+\omega^2}}.
\eea
For $\Delta\phi=\pi/6$ and $\omega=1$, there is $N\simeq9$. This
result is close to the experimental observation that there are about
a dozen bends per helix turn \cite{ZhangXB}.

According to the above analysis, to investigate the holistic nature
features of SWNTs, the in-plane strain energy should be neglected.
Thus the total energy density for SWNTs can be written as
\bea\label{E-D-D}
    \mathcal {F}_t=\frac{\pi k_c}{r_0\sqrt{1-r_0^2K^2}}+\lambda,
\eea
where $\lambda$ is the Lagrange multiplier. As to SWNTs, $\lambda$
can be taken as the average intensity of the effect between a SWNT
and it's neighbor, such as the dense SWNT ropes model in
\cite{ZhangSL2}. For MWNTs, we can take $\lambda$ as the line
tension coefficient due to the effect between one layer of MWNTs and
its two neighbor layers. Thus the above energy density in
(\ref{E-D-D}) also is suitable to each layers of MWNTs. Comparing
Eqs.~(\ref{E-M}) with (\ref{E-D-D}), we have $\lambda\sim2\pi r_0
g\sim-10$ eV/nm and $\frac{\lambda r_0}{\pi k_c}\sim-1$. This result
will be proved by our later calculations.

One can easily find that the model in (\ref{E-D-D}) will reduce to
the string model when $r_0^2K^2\ll1$. Moreover, from (\ref{E-S}) to
(\ref{E-C}), we need $r_0^2K^2<1$ not $r_0^2K^2\ll1$. This small
difference will make the model in (\ref{E-D-D}) is suitable for the
CNTs shapes with lager deformations and give us much more abundant
shapes than the string model. For a 1D elastica structure with the
energy density functional $\mathcal {F}=\mathcal {F}(K,\tau)$, the
equilibrium shape equations have been obtained in
\cite{Capovilla,Thamwattana} by discussing the first variation of
the energy $\delta(\oint\mathcal {F} ds)=0$. Making use of their
results, we attain the equilibrium shape equations for the energy
density in (\ref{E-D-D})

\bea
\nonumber
 \label{Eqs-K-1} r_0^2\big(1+r_0^2K^2-2r_0^4K^4\big)\ddot{K}+3r_0^4K\big(3+2r_0^2K^2\big)\dot{K}^2& &\\ \nonumber
                    -K\big(1-r_0^2K^2\big)^2\big(1-2r_0^2K^2+r_0^2\tau^2\big)& &\\
                      -\bar{\lambda}K(1-r_0^2K^2)^{7/2} = 0, \\
  \label{Eqs-K-2}2\dot{K}\tau(1+2r_0^2K^2)+K\dot{\tau}(1-r_0^2K^2)=0,
\eea
where we define $\bar{\lambda}=\frac{\lambda r_0}{\pi k_c}$. A ring
solution with the radius $R$ yields
\bea\label{Eq-D-R}
   (1-2r_0^2/R^2)-\bar{\lambda}(1-r_0^2/R^2)^{3/2}=0.
\eea
According to the experimental results, in most cases there are $0.5$
nm $<r_0<$ $5$ nm and $100$ nm $<R<500$ nm \cite{Martel, Cohen}, so
we have $r_0^2/R^2\rightarrow0$, which means $\bar{\lambda}\sim1$ in
Eq.~(\ref{Eq-D-R}). Specially, when $\lambda=0$ in
Eq.~(\ref{Eq-D-R}), we obtain
\bea\label{S-Clifford}
   R=\sqrt{2}r_0.
\eea
This Clifford torus solution for vesicle was found by Ou-Yang
 \cite{OuYang3} and proved by a coming experiment \cite{Mutz}. Avron and
 Berger \cite{Avron} gave some details about the torus nearby $R/r_0=2$. So the Clifford
torus with
 $R/r_0=\sqrt2$ is easy to be constructed, such as the shapes in series II of Fig.~3 of
\cite{Chuang}, which are close to this shape.

\begin{figure}\centering
\includegraphics[scale =0.8]{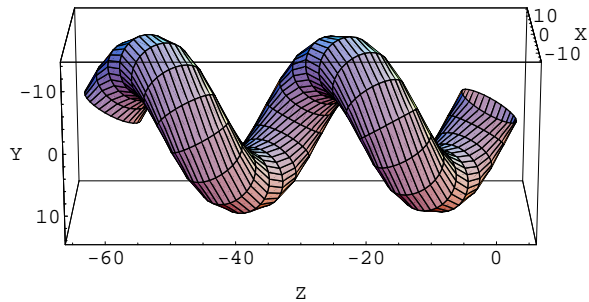}
\caption{\label{fig3} A helical shape with $R=9$ nm, $r_0=5$ nm,
pitch $H_p=2\pi h=30$ nm and $\bar{\lambda}=-0.938$ in
Eq.~(\ref{S-Helix}). This shape is in good agreement with the
experimental shape in Fig.~3(b) of \cite{Ivanov}.}
\end{figure}

Substituting the helix solution
$K=K_0\equiv\frac{R}{R^2+h^2},\tau=\tau_0\equiv\frac{h}{R^2+h^2}$
into Eq.~(\ref{Eqs-K-1}) and defining $\omega=h/R$, $\eta=r_0/R$ and
the helical angle $\psi_h=\arg\tan\omega$, we have
\bea\label{S-Helix}
\nonumber  \big(1+\omega^2\big)\big[\big(1+\omega^2\big)^2+\big(\omega^2-2\big)\eta^2\big]\\
   +\bar{\lambda} \big[\big(1+\omega^2\big)^2-\eta^2\big]^{3/2}=0.
\eea
We show an example solution in Fig.~\ref{fig3}, which is consistent
with the values in Fig.~3(b) of \cite{Ivanov}.
 Specially, the zero energy state $\omega=1$
($\psi_h=45^\circ$) \cite{ZhouXH1} yields
$\bar{\lambda}=-2/\sqrt{4-\eta^2}$. If $\lambda=0$,
Eq.~(\ref{S-Helix}) is reduced to
\bea\label{S-Helix-0}
   \eta^2=\frac{(1+\omega^2)^2}{2-\omega^2}.
\eea
Considering  $0<\eta<1$, this equation indicates
$\frac{\sqrt{2}}{2}\leq\eta<1$, $\omega^2<(\sqrt{13}-3)/2\simeq0.3$
and the helix angle $\psi_h<28.8^\circ$. However, when
$\frac{\sqrt{2}}{2}<\eta\leq0.764$, the shapes are self-intersected.
The valid region is $0.764<\eta<1$ ($14.1^\circ<\psi_h<28.8^\circ$)
and these shapes are close to the $C_{1080}$ shape in Fig.~1(b) of
\cite{Ihara1}.

\begin{figure}\centering
\includegraphics[scale =0.65]{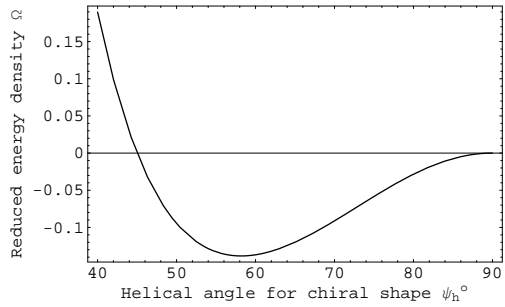}
\caption{\label{fig4} The chart of the reduced total energy
$\Omega=\Omega(\psi_h)$ in (\ref{armchair}) with $\eta=1$ for chiral
shapes. It indicates that the chiral angle difference between two
equilibrium states is $\Delta\psi_h=31.8^\circ\approx\pi/6$, which
is consistent with the lattice model.}
\end{figure}

Particularly, if $\eta=1$ and $\omega\rightarrow\infty$ in
(\ref{S-Helix}), we obtain a cylinder solution. However, if $\eta=1$
but $\omega$ is finite, what kind of shape we can obtain? If so, we
will see that it is nothing but the chiral configuration. Supposing
that the state with $\eta=1$ and $\omega\rightarrow\infty$ is the
zigzag shape which has two $sp^2$ bonds of each carbon hexagon
paralleled to the axial line of the SWNT and has the chiral angle
$\psi_h=\pi/2$, so the armchair state should be with $\psi_h=\pi/3$
 \cite{ZhangSL3} (Note that the chiral angle $\psi_h$ in this paper
is the complement angle for the definiens in \cite{ZhangSL3}). Here
we define the reduced total energy density for equilibrium helical
shapes: $\Omega= \frac{r_0}{\pi k_c}\mathcal {F}_{t}^0$, using the
method in \cite{ZhouXH1}, there is
\bea\label{armchair}
\nonumber \Omega&=&\frac{r_0}{\pi k_c}\left[\left(K^2_0-\tau^2_0\right)\mathcal {F}^0_{1}+2K_0\tau_0\mathcal {F}^0_{2}\right]/K_0\\
          &=&\frac{(1-\omega^4)\eta^2}{[(1+\omega^2)^2-\eta^2]^{3/2}},
\eea
where $\mathcal {F}_{1}=\frac{\partial\mathcal {F}_t}{\partial K}$,
$\mathcal {F}_{2}=\frac{\partial\mathcal {F}_t}{\partial \tau}$, and
 $\mathcal {F}^0=\mathcal {F}|_{K=K_0,\tau=\tau_0}$. For
straight shapes with $\eta=1$, equilibrium condition
$\partial\Omega/\partial \omega=0$ yields
\bea\label{zigzag}
\omega\rightarrow\infty,~~\omega=\sqrt{\frac{1}{3}(10^{2/3}+10^{1/3}+1)}.
\eea
These correspond to $\psi_h=\pi/2$ and
$\psi_h=58.2^\circ\approx\pi/3$, respectively. We show
$\Omega=\Omega(\psi_h)$ in Fig.~\ref{fig4} which indicates that the
chiral angle difference between two equilibrium states is about
$\pi/6$. In fact, the 2D continuous model in (\ref{E-D-D}) can not
tell us which state in (\ref{zigzag}) is the zigzag shape. However,
it gives us the angle difference between two equilibrium states.
This angle difference $\Delta\psi_h=31.8^\circ\approx\pi/6$ is in
good agreement with the lattice model. Further, considering that
each carbon hexagon has the area $s_0=3\sqrt{3}d^2/2$ and possesses
two carbon atoms, where $d=1.42$ {\AA} is the equilibrium distance
of two neighbor carbon atoms, the energy for each carbon atom is
\bea\label{atom}
E= \frac{\mathcal {F}_t^0}{4\pi r_0} s_0=\frac{3\sqrt{3}k_c
d^2}{8r_0^2}\frac{1+\omega^2}{\sqrt{(1+\omega^2)^2-\eta^2}},
\eea
where the unit for $r_0$ is \AA~and $\lambda=0$. Then choosing
$k_c=1.62$ eV, \cite{TuZC2008} we have $E=\frac{2.1217}{r_0^2}$ and
$E=\frac{2.20869}{r_0^2}$ for the two equilibrium states in
(\ref{zigzag}) (note $\eta=1$), respectively. We simulated the
corresponding values in \cite{Adams} obtained by QMD method and
found that the energy for each carbon atoms satisfies
$E=\frac{2.05765}{r_0^2}$ for zigzag shapes and
$E=\frac{2.08964}{r_0^2}$ for armchair shapes. Clearly, the result
obtained by the continuous method in (\ref{atom}) is close to the
corresponding result obtained by QMD method, and it also indicates
the two states in (\ref{zigzag}) are the zigzag shape and the
armchair shape, respectively. If we want that the result in
(\ref{atom}) can consistent with the QMD result in \cite{Adams} more
exactly, we can choose $k_c=1.55$ eV. Moreover, for a chiral shape
with $\eta=1$ and $\omega\neq0$, and a helical shape with $\eta<1$
and the same $\omega$ as the chiral shape, Eq.~(\ref{atom})
indicates the helical shape will have lower energy for each atom. It
explains why there are so many helical shapes are found in
experiments. The above results indicate that the continuous model in
(\ref{E-D-D}) which based on taking the $sp^{2}$ bonds as the
geodesic lines on a tube also contains a little information about
the lattice structure.

\subsection{Numerical results}
\begin{figure}\centering
\includegraphics[scale =0.7]{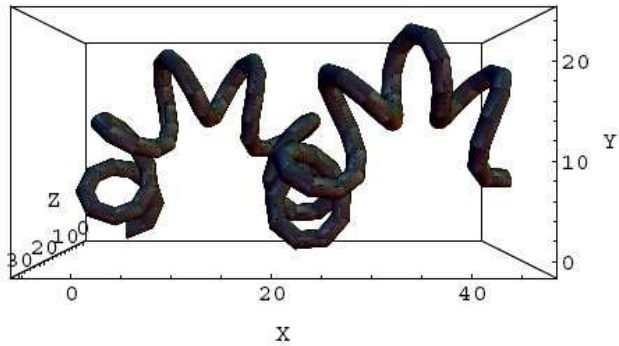}
\caption{\label{fig5} A positive super helical shape with the
initial conditions: $r_0=1$ nm, $\bar{\lambda}=-0.978$,
$\theta(0)=\varphi(0)=\ddot{\varphi}(0)=0$, $\dot{\theta}(0)=0.04$
nm$^{-1}$, $\dot{\varphi}(0)=0.26$ nm$^{-1}$ and
$\ddot{\theta}(0)=0.01$ nm$^{-2}$. The first helix is right-handed
and the second helix is left-handed. }
\end{figure}

\begin{figure}\centering
\includegraphics[scale =0.7]{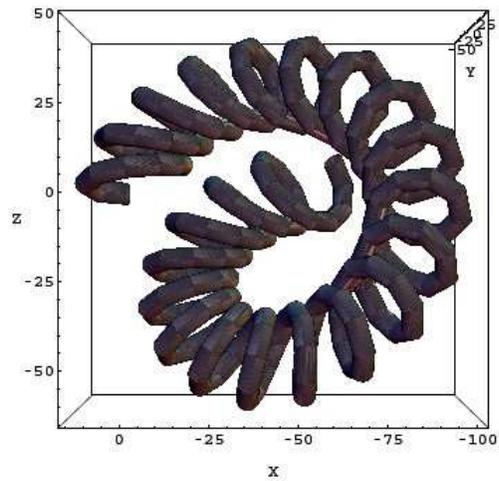}
\caption{\label{fig6} A negative super helical shape with the
initial conditions: $r_0=3$ nm, $\bar{\lambda}=-0.97$,
$\theta(0)=\varphi(0)=0$, $\dot{\theta}(0)=0.018$ nm$^{-1}$,
$\dot{\varphi}(0)=0.08$ nm$^{-1}$, $\ddot{\theta}(0)=0.0009$
nm$^{-2}$ and $\ddot{\varphi}(0)=0.0011$ nm$^{-2}$. The first and
the second helixes are right-handed.}
\end{figure}

\begin{figure}\centering
\includegraphics[scale =0.6]{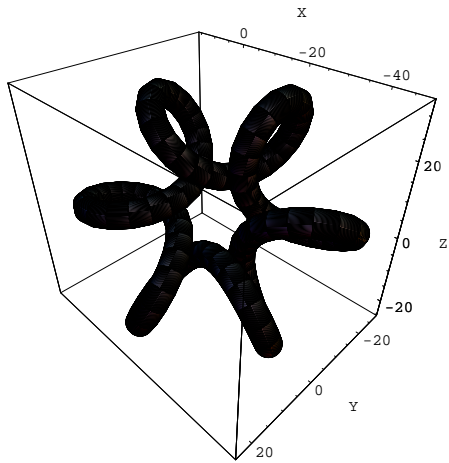}
\caption{\label{fig7} A right-handed helical ring with the initial
conditions: $r_0=3$ nm, $\bar{\lambda}=-0.96$,
$\theta(0)=\varphi(0)=0$, $\dot{\theta}(0)=0.002$ nm$^{-1}$,
$\dot{\varphi}(0)=0.07$ nm$^{-1}$, $\ddot{\theta}(0)=0.002$
nm$^{-2}$ and $\ddot{\varphi}(0)=0.00261$ nm$^{-2}$.}
\end{figure}

Let the central line of the SWNT be $\textbf{R} = \{R_x,R_y,R_z\}$,
$\dot{R_x}=\cos\theta\cos\varphi$, $\dot{R_y}=\cos\theta\sin\varphi$
and $\dot{R_z}=\sin\theta$, where $\theta=\theta(s)$ and
$\varphi=\varphi(s)$ are two Euler angles with the variable $s$,
 there are
\bea
 \label{Euler-1}      K^2 &=& \dot{\theta}^2+\dot{\varphi}^2\cos^2\theta,\\
 \label{Euler-2}    \tau K^2&=&\dot{\varphi}(2\dot{\theta}^2+\dot{\varphi}^2\cos^2\theta)\sin\theta
   +(\dot{\varphi}\ddot{\theta}-\dot{\theta}\ddot{\varphi})\cos\theta.~~~~~
\eea
Substituting the above expressions into Eqs.~(\ref{Eqs-K-1}) and
(\ref{Eqs-K-2}), we obtain two tedious third order equations about
$\theta$ and $\varphi$ (see appendix \ref{appendix-B}). Solving this
two equations, we obtain several interesting shapes. Fig.~\ref{fig5}
shows a positive super helical tube, Fig.~\ref{fig6} shows a
negative super helix tube and Fig.~\ref{fig7} shows a right handed
helical ring. Experimental basketwork in Fig.~2(b) of \cite{ZhangM3}
which contains many super helical MWNT shapes has excellent
mechanical capabilities. As to these unattached super helical shapes
in Fig.~\ref{fig5} and Fig.~\ref{fig6}, we can conclude that they
have strong and restorable retractility like the DNA chain. So, they
are perfect functional materials and have large potential in
constructing nano-instruments.

In the planar case, the valid shape equation (\ref{Eqs-K-1}) is
reduced to
\bea\label{Eq-Planar}
\nonumber r_0^2\big(1+\Upsilon^2-2\Upsilon^4\big)\ddot{\Upsilon}+3r_0^2\Upsilon\big(3+2\Upsilon^2\big)\dot{\Upsilon}^2\\
                             -\Upsilon\big(1-\Upsilon^2\big)^2\big(1-2\Upsilon^2\big)-\bar{\lambda}\Upsilon\big(1-\Upsilon^2\big)^{7/2}
                             =0,
\eea
where $\Upsilon=r_0K$. This is a non-linear equation and it is
difficult to be solved generally. We numerically solved this
equation but all the solutions we obtained are similar to the shapes
in Fig.~4 of \cite{ZhouXH3}.

\section{Conclusions}\label{Conclusions}
 In conclusion, we have shown a connection between the elastic shell model
 and the Kirchhoff elastic rod model. Combining
Refs.~\cite{OuYang1,TuZC} and this work, a complete method to deal
with CNTs has been constructed. From lattices model to 2D continues
elastic shell model and further to 1D continues elastic rod model,
this method gives a recommendable approach to del with the
multi-scale low dimensional systems. The 1D model in
(\ref{E-D-Tube-2}) which contains the in-plane strain energy is
suitable to investigate the mechanical behaviors of CNTs. But we
should note that the first term on the right hand side of
(\ref{E-D-Tube-2}) is the distinct difference, when compared with
the isotropic Kirchhoff elastic rod model. This difference will make
CNTs have unusual mechanical behaviors which have not been known.
Moreover, we should note that the model in (\ref{E-D-SWNT}) is
obtained by taken CNTs as isotropic perfect tubes. If there are
plenty of pentagonal and heptagonal defects for bended CNTs, this
model seams need to be changed. A simple way to adapt this change is
to adjust the constants $k_c$ and $k_d$, such as the work in
\cite{Chopra} where the authors chose $k_c=1.4$ eV which is
different to the value $k_c=1.62$ eV in \cite{TuZC2008}.

 For straight chiral CNTs, our study indicates that the difference of the
chiral angle between two equilibrium states is about $\pi/6$, which
is consistent with the angle difference between the zigzag shape and
the armchair shape obtained by lattice model. Our study also reveals
that, if a helical shape and a straight chiral shape have the same
radius $r_0$ and the same $\omega$, the former structure will have
lower energy for per atom than the later one, which explains why
there are so many helical CNTs in experiments. Since there are super
helical solutions for the equilibrium shape equations, we hope they
can be found in future experiments and the super retractility also
can be found for these shapes (If it is only the string model
\cite{OuYang1}, there are only helical solutions but not super
helical shapes \cite{Langer}). How to produce super helical CNTs in
experiments? Yin \emph{et al.} \cite{Yin} provided a way to
construct the super CNTs using a cylindrical template. If we choose
a helical template, super helical CNTs may be available.

Finally, we would like to point out that how to use the lattice
model to construct the super helical CNTs needs further discussion,
which will be our future work.

\section*{Acknowledgements}
The author would like to thank Zhanchun Tu, Weihua Mu and Jianlin
Liu for helpful suggestions.

\appendix{}
\section{}\label{appendix-A}
Defining $(\mu,\nu)$ to be the local coordinates on the tube
$\textbf{Y}$, where $\mu$ is along the tangent of the central line
of the tube $\textbf{R}$ and $\nu$ is upright to $\textbf{R}$, there
are $d\mu=ds$ and $d\nu=r_0d\theta$. Making use of the Frenet
formulae, we obtain
\bea\label{A1}
E&=&\textbf{Y}_\mu\cdot\textbf{Y}_\mu=(1-r_0K\cos\theta)^2+r_0^2\tau^2,\\
F&=&\textbf{Y}_\mu\cdot\textbf{Y}_\nu=r_0\tau,\\
G&=&\textbf{Y}_\nu\cdot\textbf{Y}_\nu=1,
\eea
where $\textbf{Y}_\mu=\partial\textbf{Y}/\partial\mu$. The main
normal vector of the tube is
\bea\label{A2}
\textbf{n}=\frac{\textbf{Y}_\mu\times\textbf{Y}_\nu}{|\textbf{Y}_\mu\times\textbf{Y}_\nu|}=\textbf{r}/r_0.
\eea
Consequently we obtain
\bea\label{A3}
L&=&\textbf{Y}_{\mu\mu}\cdot\textbf{n}=K\cos\theta(1-r_0K\cos\theta)^2-r_0\tau^2,\\
M&=&\textbf{Y}_{\mu\nu}\cdot\textbf{n}=-\tau,\\
N&=&\textbf{Y}_{\nu\nu}\cdot\textbf{n}=-1/r_0.
\eea
The mean curvature and Gaussian curvature are
\bea\label{A4}
H&=&\frac{LG-2MF+NE}{2(EG-F^2)}=\frac{1-2r_0K\cos\theta}{2r_0(r_0K\cos\theta-1)},\\
\Lambda&=&\frac{LN-M^2}{EG-F^2}=\frac{K\cos\theta}{r_0(r_0K\cos\theta-1)}.
\eea

\section{}\label{appendix-B}

In this part, we show the shape equations (\ref{Eqs-K-1}) and
(\ref{Eqs-K-2}) with the Euler angles as variables. Substituting
expressions (\ref{Euler-1}) and (\ref{Euler-2}) into
Eqs.~(\ref{Eqs-K-1}) and (\ref{Eqs-K-2}), we obtain two tedious
equations
\bea\label{Eqs-Euler-1}
 \nonumber
 4r_0^2\big(\dot{\theta}\dot{\varphi}^2\sin\theta\cos\theta-\dot{\theta}\ddot{\theta}-\dot{\varphi}\ddot{\varphi}\cos^2\theta\big)^2\\\nonumber
       \times\Big\{8r_0^2(\dot{\theta}^2+\dot{\varphi}^2\cos^2\theta)\big[1+r_0^2(\dot{\theta}^2+\dot{\varphi}^2\cos^2\theta)\big]-1\Big\}\\\nonumber
     +4r_0^2\Big\{1+r_0^2(\dot{\theta}^2+\dot{\varphi}^2\cos^2\theta)
               \big[1-2r_0^2(\dot{\theta}^2+\dot{\varphi}^2\cos^2\theta)\big]\Big\}\\\nonumber
               \times(\dot{\theta}^2+\dot{\varphi}^2\cos^2\theta)\big(\ddot{\theta}^2+\ddot{\varphi}^2\cos^2\theta
               -\dot{\theta}^2\dot{\varphi}^2\cos2\theta\\\nonumber
               -\ddot{\theta}\dot{\varphi}^2\sin\theta\cos\theta-2\dot{\theta}\dot{\varphi}\ddot{\varphi}\sin2\theta+\dot{\theta}\theta^{(3)}+\dot{\varphi}\varphi^{(3)}\cos^2\theta\big)\\\nonumber
               -4\big[1-r_0^2(\dot{\theta}^2+\dot{\varphi}^2\cos^2\theta)\big]^2 \times\Big\{(\dot{\theta}^2+\dot{\varphi}^2\cos^2\theta)^2\\\nonumber
               \times\Big[1-2r_0^2(\dot{\theta}^2+\dot{\varphi}^2\cos^2\theta)
               +\bar{\lambda}\big(1-r_0^2(\dot{\theta}^2+\dot{\varphi}^2\cos^2\theta)\big)^{3/2}\Big]\\\nonumber
               +r_0^2\big[2\dot{\theta}^2\dot{\varphi}\sin\theta
               +\dot{\varphi}\cos\theta(\dot{\varphi}^2\sin\theta\cos\theta+\ddot{\theta})
               -\dot{\theta}\ddot{\varphi}\cos\theta\big]^2\Big\}\\
               =0,~~~~
\eea
\bea\label{Eqs-Euler-2}
\nonumber
      \big(\dot{\theta}\ddot{\theta}+\dot{\varphi}\ddot{\varphi}\cos^2\theta-\dot{\theta}\dot{\varphi}^2\sin\theta\cos\theta\big)
          \big[1+2r_0^2(\dot{\theta}^2+\dot{\varphi}^2\cos^2\theta)\big]\\\nonumber
           \times2\big[2\dot{\theta}^2\dot{\varphi}\sin\theta+\dot{\varphi}\cos\theta(\dot{\varphi}^2\sin\theta\cos\theta+\ddot{\theta})-\dot{\theta}\ddot{\varphi}\cos\theta\big]\\\nonumber
          +\big[1-r_0^2(\dot{\theta}^2+\dot{\varphi}^2\cos^2\theta)\big]\Big\{2\dot{\theta}^5\dot{\varphi}\cos\theta
               +3\dot{\theta}^4\ddot{\varphi}\sin\theta\\\nonumber
               +\dot{\varphi}^2\cos^3\theta\big(\dot{\varphi}^2\ddot{\varphi}\sin\theta\cos\theta-2\ddot{\theta}\ddot{\varphi}+\dot{\varphi}\theta^{(3)}\big)\\\nonumber
               +\dot{\theta}^2\cos\theta(2\ddot{\theta}\ddot{\varphi}+\dot{\varphi}\theta^{(3)})
               +\frac{1}{4}\dot{\theta}^3\big[\dot{\varphi}^3(11\cos\theta+\cos3\theta)\\\nonumber
               -4\ddot{\theta}\dot{\varphi}\sin\theta
               -4\varphi^{(3)}\cos\theta\big]
               +\dot{\varphi}\dot{\theta}\cos\theta\big[3\dot{\varphi}^2\ddot{\theta}\sin\theta\cos\theta
               -2\ddot{\theta}^2\\\nonumber
               +\cos^2\theta(\dot{\varphi}^4\cos^2\theta
               +2\ddot{\varphi}^2-\dot{\varphi}\varphi^{(3)})\big]\Big\}\\
               =0.~~~~
\eea
Solving the above equations, we get $\theta=\theta(s)$ and
$\varphi=\varphi(s)$, and consequently obtain $R_x=\int_0^S
\cos\theta\cos\varphi ds$, $R_y=\int_0^S \cos\theta\sin\varphi ds$
and $R_z=\int_0^S\sin\theta ds$. The SWNT shape can be written as
\bea\label{Y-Shape}
   \nonumber&&\textbf{Y} = \{X,Y,Z\},\\\nonumber
            &&X=(\dot{\theta}^2+\dot{\varphi}^2\cos^2\theta)^{\frac{-1}{2}}\big[r_0\sin\varphi(\dot{\theta}\sin\phi-\dot{\varphi}\cos\theta\cos{\phi})\\
            &&~~~~~-r_0\sin\theta\cos\varphi(\dot{\theta}\cos\phi+\dot{\varphi}\cos\theta\sin{\phi})\big]+R_x,\\\nonumber
            &&Y=(\dot{\theta}^2+\dot{\varphi}^2\cos^2\theta)^{\frac{-1}{2}}\big[r_0\cos\varphi(\dot{\varphi}\cos\theta\cos\phi-\dot{\theta}\sin{\phi})\\
            &&~~~~~-r_0\sin\theta\sin\varphi(\dot{\theta}\cos\phi+\dot{\varphi}\cos\theta\sin{\phi})\big]+R_y,\\\nonumber
            &&Z=r_0\cos\theta(\dot{\theta}^2+\dot{\varphi}^2\cos^2\theta)^{\frac{-1}{2}}(\dot{\theta}\cos\phi+\dot{\varphi}\cos\theta\sin\phi)\\
            &&~~~~~+R_z.
\eea

\end{document}